\documentclass[prd, superscriptaddress,nofootinbib]{revtex4}
\usepackage{graphicx}
\usepackage{dcolumn}
\usepackage{slashbox,multirow}
\usepackage{amsmath,amsthm,amssymb}
\begin{document}
%


\title{Late-time tails of wave maps coupled to gravity}

\author{Piotr Bizo\'n}
\affiliation{M. Smoluchowski Institute of Physics, Jagiellonian
University, Krak\'ow, Poland}
\author{Tadeusz Chmaj}
\affiliation{H. Niewodniczanski Institute of Nuclear
   Physics, Polish Academy of Sciences,  Krak\'ow, Poland}
   \affiliation{Cracow University of Technology, Krak\'ow,
    Poland}
\author{Andrzej Rostworowski}
\affiliation{M. Smoluchowski Institute of Physics, Jagiellonian
University, Krak\'ow, Poland}
\author{Stanis{\l}aw Zaj\c{a}c}
\affiliation{H. Niewodniczanski Institute of Nuclear
   Physics, Polish Academy of Sciences,  Krak\'ow, Poland}
\date{\today}
\begin{abstract}
We consider the late-time asymptotic behavior for solutions of Einstein's equations with the wave
map matter. Solutions starting from small compactly supported $\ell$-equivariant initial data
with $\ell\geq 1$ are shown to decay as $t^{-(2\ell+2)}$ at future timelike infinity and as
$u^{-(\ell+1)}$ at future null infinity.

\end{abstract}

\maketitle

\section{Introduction}
In this paper we continue our investigations, initiated in \cite{bcr5}, of the precise
\emph{quantitative} description of the late-time asymptotic behavior of self-gravitating massless
fields. In \cite{bcr5} we considered the simplest case of a spherically symmetric massless scalar
field. Using nonlinear perturbation analysis we showed that solutions  starting from small
initial data decay as $t^{-3}$ at timelike infinity and as $u^{-2}$ at null infinity.  We also
derived a simple analytic formula for the amplitude of the late-time tail in terms of initial
data.

Here we study the analogous problem for wave maps which are a natural geometric generalization of
the wave equation for the massless scalar field. This generalization seems interesting because in
the so called equivariant case the homotopy index $\ell$ of the map plays the role similar to the
multipole index for spherical harmonics. However, in contrast to the decomposition of a scalar
field into spherical harmonics which makes sense only at the linearized level, it is consistent
to study nonlinear evolution for the wave map within a fixed equivariance class. In this sense
$\ell$-equivariant self-gravitating wave maps can serve as a poor man's toy-model of
non-spherical collapse. The $\ell=0$ case reduces to the spherically symmetric massless scalar
field analyzed in \cite{bcr5} so hereafter we assume that $\ell\geq 1$. We note aside that the
$\ell=1$ case has been extensively studied in the past focusing on the critical behavior at the
threshold of black hole formation \cite{bw1, h, bw2, l, abt}, however, to our knowledge, the
late-time behavior of wave maps coupled to gravity has not been analyzed before.

Using the same third-order perturbation method as in \cite{bcr5} we show here that for small
compactly supported initial data the late-time tail of the self-gravitating $\ell$-equivariant
wave map decays as $t^{-(2\ell+2)}$ at future timelike infinity and as $u^{-(\ell+1)}$ at future
null infinity. We also compute the amplitude of the tail in terms of initial data. These analytic
results are verified by the numerical integration of the Einstein-wave map equations.
\section{Setup}
Let $U: \mathcal{M} \rightarrow \mathcal{N}$ be a map from a spacetime $(\mathcal{M},g_{ab})$
into a Riemannian manifold $(\mathcal{N},G_{AB})$. A pair $(U,g_{ab}$) is said to be a wave map
coupled to gravity if it is a critical point of the action functional
\begin{equation}\label{action}
S = \int_{\mathcal{M}} \left(\frac{R}{16 \pi G} -\frac{\lambda}{2} g^{ab} \partial_a U^A
\partial_b U^B
 G_{AB}\right) dv \,,
\end{equation}
where $R$ is the scalar curvature of the metric $g_{ab}$,  $G$ is Newton's constant, $\lambda$ is
the wave map coupling constant, and $dv$ is the volume element on $(\mathcal{M},g_{ab})$. The
field equations derived from (\ref{action}) are the Einstein equations $R_{ab}-\frac{1}{2} g_{ab}
R = 8 \pi G T_{ab}$ with the stress-energy tensor
\begin{equation}\label{tep}
 T_{ab} = \lambda \left(\partial_a U^A \partial_b U^B
 -\frac{1}{2} g_{ab}( g^{cd} \partial_c U^A \partial_d U^B)\right)
 G_{AB}\,,
\end{equation}
and the wave map equation
\begin{equation}\label{weq}
\square_g U^A + \Gamma_{BC}^A(U) \partial_a U^B \partial_b U^C g^{ab}=0,
\end{equation}
where $\Gamma_{BC}^A$ are the Christoffel symbols of the target metric $G_{AB}$ and $\square_g$
is the wave operator associated with the metric $g_{ab}$. As a target manifold we take the
three-sphere with the round metric in polar coordinates $U^A=(F,\Omega)$
\begin{equation}\label{metric_t}
G_{AB} dU^A dU^B = dF^2 + \sin^2{\!F} \:d\Omega^2.
\end{equation}
For the four dimensional spacetime $\mathcal{M}$ we assume spherical symmetry and use the
following ansatz for the metric
\begin{equation}\label{metric_d}
g_{ab}dx^a dx^b = e^{2\alpha(t,r)}\left(-e^{2\beta(t,r)} dt^2 + dr^2\right) + r^2 d\omega^2\,.
\end{equation}
In addition we assume that the map $U$ is spherically $\ell$-equivariant, that is
\begin{equation}\label{equivariance}
F=F(t,r),\qquad \Omega=\chi_{\ell}(\omega)\,,
\end{equation}
where $\chi_{\ell}(\omega)$ is a homogeneous harmonic  polynomial of degree $\ell$. For this
ansatz, the energy-momentum tensor (\ref{tep}) does not depend on angles and thus can be
consistently coupled to the spherically symmetric Einstein equations. We note in passing that a
very similar idea of introducing the "angular momentum" into spherical collapse was put forward
by Olabarrieta \emph{et al.} \cite{ovcu} in the context of critical phenomena.
 In terms of the mass function $m(t,r)=\dfrac{1}{2}r(1-e^{-2\alpha})$ the Einstein
equations take the following form (hereafter primes and dots denote partial derivatives with
respect to $r$ and $t$, respectively)
\begin{eqnarray}
\label{h-constraint} m'&=& \frac{\kappa}{2}\, r^2 e^{-2\alpha} \left(F'^2 + e^{-2\beta} \dot
F^2\right)+\kappa \frac{\ell(\ell+1)}{2} \sin^2{\!F}\,,
\\
\dot{m}&=& \kappa \,r^2 e^{-2\alpha} \dot F \,F'\,,
\\
\label{s-condition} \beta' &=& \frac{2m}{r^2} e^{2\alpha}-\kappa \ell(\ell+1)\, e^{2\alpha}
\frac{\sin^2{\!F}}{r}\,,
\end{eqnarray}
where $\kappa=4\pi G \lambda$ is a dimensionless parameter.
 The
wave map equation (\ref{weq}) takes the form
\begin{equation}\label{we2}
    \left(e^{-\beta}\dot F\right)^{\cdot}-\frac{1}{r^2}\left(r^2 e^{\beta}
    F'\right)'+e^{\beta+2\alpha}\ell(\ell+1)\frac{\sin{
    2F}}{2r^2}=0\,.
\end{equation}
For $\ell=0$ the above equations reduce to the Einstein-massless scalar field equations analyzed
by us in \cite{bcr5}.
 For $\kappa=0$ (no gravity) equations (7)-(9) are trivially solved by $m=0$ and
 $\beta=0$, while equation (\ref{we2}) reduces to the flat space wave map equation.
\section{Iterative scheme}
\label{sec:eqs}

We assume that initial data are small, smooth, and compactly supported (the last assumption can
be replaced by a suitable fall-off condition)
\begin{equation}
\label{id} F(0,r) = \varepsilon g(r), \qquad \dot F(0,r) = \varepsilon h(r)\,.
\end{equation}
We make the following perturbative expansion
\begin{eqnarray}
m(t,r) &=& m_0(t,r) + \varepsilon m_1(t,r) + \varepsilon^2 m_2(t,r) + \dots,
\\
\beta(t,r) &=& \beta_0(t,r) + \varepsilon \beta_1(t,r) + \varepsilon^2 \beta_2(t,r) + \dots,
\\
F(t,r) &=& F_0(t,r) + \varepsilon F_1(t,r) + \varepsilon^2 F_2(t,r) + \varepsilon^3 F_3(t,r) +
\dots. \label{pertexpansion}
\end{eqnarray}
Substituting this expansion into the field equations and grouping terms with the same power of
$\varepsilon$ we get the iterative scheme which can be solved recursively.

We consider perturbations about Minkowski spacetime, so $m_0=\beta_0=F_0=0$. At the first order
the metric functions $m_1=\beta_1=0$ (this follows from regularity at $r=0$), while $F_1$
 satisfies the flat space radial wave equation for the $\ell$-th spherical harmonic
\begin{equation}
\label{Box_f1} \Box_{(\ell)} F_1 =0\,,\qquad
\Box_{(\ell)}=\partial_t^2-\partial_r^2-\frac{2}{r}\partial_r+\frac{\ell(\ell+1)}{r^2}\,,
\end{equation}
with initial data $F_1(0,r) = g(r), \dot F_1(0,r) =  h(r)$.
The general everywhere  regular solution of equation (\ref{Box_f1}) is given by a superposition
of outgoing and ingoing waves
\begin{equation}
\label{f1} F_1 (t,r) = F_1^{ret}(t,r) + F_1^{adv}(t,r)\,,
\end{equation}
where
\begin{equation}
\label{f1retadv} F_1^{ret}(t,r) = \frac{1}{r}\,\sum_{k=0}^{l}  \frac {(2\ell-k)!} {k!(\ell-k)!}
\frac {a^{(k)}(u)}{(v-u)^{\ell-k}}\,, \qquad   F_1^{adv}(t,r) = \frac{1}{r}\,\sum_{k=0} ^{\ell}
(-1)^{k+1} \frac {(2\ell-k)!} {k!(\ell-k)!} \frac {a^{(k)}(v)}{(v-u)^{\ell-k}}\,,
\end{equation}
and $u=t-r$, $v=t+r$ are the retarded and advanced times, respectively (the superscript in round
brackets denotes the $k$-th derivative). Note that for compactly supported initial data the
generating function $a(x)$ can be chosen to have compact support as well (this condition
determines $a(x)$ uniquely).

At the second order $\Box_{(\ell)} F_2 = 0$, hence $F_2=0$ (because it has zero initial data),
while the metric functions satisfy the following equations
\begin{eqnarray}
\label{m2prime} m'_2 &=& \frac{\kappa}{2}\, r^2 \left( \dot{F}_1^2 + F_1'^2 +
\frac{\ell(\ell+1)}{r^2} F_1^2\right)\,,
\\
\label{m2dot} \dot{m}_2 &=&  \kappa\, r^2 \dot{F}_1 F'_1\,,
\\
\label{beta2prime} \beta'_2 &=& \frac{2 m_2}{r^2} - \kappa \frac{\ell(\ell+1)}{r} F_1^2\,.
\end{eqnarray}
We temporarily postpone the discussion of this system and proceed now to the third order, where
we have
\begin{equation}
\label{Box_f3} \Box_{(l)} F_3  = 2 \beta_2 \ddot{F}_1 + \dot{\beta}_2 \dot{F}_1+ \beta'_2 F'_1 -
\frac{2\ell(\ell+1) m_2 F_1}{r^3} + \frac{2\ell(\ell+1) F_1^3}{3 r^2}\,.
\end{equation}
To solve this equation we use the Duhamel formula for the solution of the inhomogeneous wave
equation $\Box_{(\ell)} F=N(t,r)$ with zero initial data
\begin{equation}
\label{duh} F(t,r)= \frac {1} {2 r} \int \limits_{0}^{t} d\tau \int
\limits_{|t-r-\tau|}^{t+r-\tau} \rho P_{\ell}(\mu)  N(\tau,\rho) d\rho\,,
\end{equation}
where $P_{\ell}(\mu)$ are Legendre polynomials of degree $\ell$  and
$\mu=(r^2+\rho^2-(t-\tau)^2)/2r\rho$ (note that $-1\leq \mu \leq 1$ within the integration
range). Applying this formula to equation (\ref{Box_f3}), using null coordinates $\eta=\tau-\rho$
and $\xi=\tau+\rho$, and the abbreviation $K(m,\beta,F)=2 \beta \ddot{F} + \dot{\beta} \dot{F}+
\beta' F' - (2\ell(\ell+1)/r^2)( m F / r - F^3 / 3)$, we obtain
\begin{equation}
\label{f3} F_3(t,r) = \frac {1} {8 r} \int\limits_{|t-r|}^{t+r} d\xi \int \limits_{-\xi}^{t-r}
(\xi-\eta) P_{\ell}(\mu) K(m_2(\xi,\eta),\beta_2(\xi,\eta),F_1(\xi,\eta)) d\eta\,,
\end{equation}
where now $ \mu=(r^2+(\xi-t)(t-\eta))/r(\xi-\eta)$. If the initial data (\ref{id}) vanish outside
a ball of radius $R$, then for $t>r+R$ we may drop the advanced part of $F_1(t,r)$ and
interchange the order of integration in (\ref{f3}) to get
\begin{equation}
\label{f3(2)} F_3(t,r) = \frac {1} {8 r} \int \limits_{-\infty}^{\infty} d\eta \int
\limits_{t-r}^{t+r} (\xi-\eta) P_{\ell}(\mu)
K(m_2(\xi,\eta),\beta_2(\xi,\eta),F^{ret}_1(\xi,\eta)) \, d\xi \,.
\end{equation}
In order to determine the late-time behavior of $F_3(t,r)$ we need to know the behavior of the
source term $K$ along the light cone for large values of $r$ (the intersection of the integration
range in (\ref{f3(2)}) with the support of $F_1^{ret}(t,r)$).
Having that, we shall expand the function $K$ in (\ref{f3(2)}) in the inverse powers of
$\rho=(\xi-\eta)/2$ and calculate the integrals using the following identity (see the appendix in
\cite{BCR4} for the derivation)
\begin{equation}\label{master}
\int \limits_{t-r}^{t+r} d\xi \, \frac {P_{\ell} (\mu)} {(\xi-\eta)^{n}} =  (-1)^l \frac
{2(n-2)^{\underline{\ell}}} {(2\ell+1)!!} \, \frac {r^{\ell+1}(t-\eta)^{n-\ell-2}} {[(t-\eta)^2 -
r^2]^{n-1}} \, F \left( \left. \begin{array}{c} \frac {\ell+2-n} {2}, \, \frac {\ell+3-n} {2}  \\
\ell + 3/2
\end{array} \right| \left( \frac {r} {t - \eta} \right)^{2} \right)\,,
\end{equation}
where $(n-2)^{\underline \ell}=(n-2)(n-3)\cdots(n-\ell-1)$.

Now, we return to the analysis of the second-order equations (18)-(20). Substituting the outgoing
solution (\ref{f1retadv}) into (\ref{m2prime}) and integrating, we get
\begin{eqnarray}
m_2(t,r) &\stackrel{t>R}{=}& \kappa \int \limits_0^r \left[ \left(a^{(\ell+1)}(t-\rho)\right)^2 - \sum_{1 \leq k \leq 2\ell+2} \; \sum_{0 \leq n \leq k-1} \frac {(\ell+n)^{\underline{2n}} (\ell+k-1-n)^{\underline{2(k-1-n)}} \left( \ell^2 + \ell + (k-n)(n+1) \right)} {k 2^{k} \, (k-1-n)! \, n!} \right.
\nonumber\\
&& \hskip 3.5cm \left. \partial_{\rho} \frac { a^{(\ell+1+n-k)}(t-\rho) \, a^{(\ell-n)}(t-\rho)} {\rho^k} \right]\, d\rho \,,
\label{m2f}
\end{eqnarray}
where we used that $m_2(t,r=0)=0$, which follows from regularity of initial data at the origin
and (\ref{m2dot}). Here and in the following we use repeatedly the fact that $a(x)=0$ for
$|x|>R$, $R$ being the radius of a ball on which the initial data (\ref{id}) are supported. To
describe the behavior of $m_2(t,r)$ along the lightcone it is convenient to use the null
coordinate $u=t-r$ instead of $t$, and rewrite (\ref{m2f}) as
\begin{equation}
m_2(u,r) \stackrel{r+u>R}{=} \kappa \left[ \int \limits_u^\infty \left(a^{(\ell+1)}(s)\right)^2
\, ds - \frac {{\ell}^2+\ell+1} {2 r} \left(a^{(\ell)}(u)\right)^2 -
\frac{\ell(\ell+1)({\ell}^2+\ell+2)}{4} \frac {a^{(\ell-1)}(u) a^{(\ell)}(u)} {r^2} + \mathcal{O} \left( \frac {1}
{r^3} \right) \right]\,.
\end{equation}
Next, using the gauge freedom to set $\beta_2(t,r=0)=0$ and integrating equation
(\ref{beta2prime}), we get
\begin{eqnarray}
\beta_2(t,r) &\stackrel{t>R}{=}& 2 \kappa \int \limits_0^r \frac {1} {\rho^2} \int \limits_{t-\rho}^{\infty} \left(a^{(\ell+1)}(s)\right)^2 \, ds \, d\rho
\nonumber\\
&-& \hskip -3mm \kappa \int \limits_0^r \left[ (2{\ell}^2+2\ell+1) \frac {\left(a^{(\ell)}(t-\rho)\right)^2} {\rho^3} + \frac{\ell(\ell+1)(3{\ell}^2+3\ell+2)}{2} \frac {a^{(\ell-1)}(t-\rho)a^{(\ell)}(t-\rho)} {\rho^4} + \mathcal{O} \left(\frac {1} {\rho^5} \right) \right] d\rho.
\end{eqnarray}
The first integral can be integrated by parts giving
\begin{eqnarray}
\beta_2(u,r) &\stackrel{r+u>R}{=}& 2 \kappa \left[ -\frac{1}{r} \int \limits_u^{\infty} \left(a^{(\ell+1)}(s)\right)^2 \, ds + \int \limits_{u}^{\infty} \frac {\left(a^{(\ell+1)}(s)\right)^2} {r-(s-u)} \, ds \right]
\nonumber\\
&-& \kappa \int \limits_{u}^{\infty} \left[ (2{\ell}^2+2\ell+1) \frac {\left(a^{(\ell)}(s)\right)^2} {(r-(s-u))^3} + \frac{\ell(\ell+1)(3{\ell}^2+3\ell+2)}{2} \frac {a^{(\ell-1)}(s)a^{(\ell)}(s)} {(r-(s-u))^4} \right] \, ds + \mathcal{O} \left( \frac {1} {r^5} \right) \, .
\label{Beta2}
\end{eqnarray}
%
\section{Tails}
Now, we shall apply the method described above to compute the late-time asymptotics of solutions
in the third-order approximation. Hereafter, it is convenient to define the following integrals
(for non-negative integers $m,n$)
\begin{equation}
\label{Imn} I^{m}_{n} (u) = \int \limits_u^\infty (s-u)^m \left(a^{(n)}(s)\right)^2 \, ds.
\end{equation}
\subsection{$\mathbf{\ell=1}$}
From (\ref{Beta2}) we have
\begin{eqnarray}
\label{beta2} \beta_2(u,r) &\stackrel{r+u>R}{=}& \frac {\kappa} {r^2} \left[ 2 I^{1}_{2}(u) +
\frac{1}{r} (2 I^{2}_{2}(u) - 5 I^{0}_{1}(u)) + \mathcal{O} \left( \frac {1} {r^2} \right)
\right]\,,
\\
\label{beta2dot} \dot{\beta}_2(u,r) &\stackrel{r+u>R}{=}& - \frac {\kappa} {r^2} \left[ 2
I^{0}_{2}(u) + \frac{1}{r} (4 I^{1}_{2}(u) - 5 \left(a'(u)\right)^2) + \mathcal{O} \left( \frac
{1} {r^2} \right) \right]\,,
\\
\label{beta2prime(2)} \beta_2'(u,r)  &\stackrel{r+u>R}{=}& \frac {\kappa} {r^2} \left[ 2
I^{0}_{2}(u) - \frac{5}{r} \left(a'(u)\right)^2 + \mathcal{O} \left( \frac {1} {r^2} \right)
\right]\,.
\end{eqnarray}
Substituting (\ref{f1retadv}) and (\ref{beta2}-\ref{beta2prime(2)}) into (\ref{f3(2)}) we obtain
\begin{eqnarray}
\label{F3(3)} F_3(t,r) &=& \frac {4 \kappa} {r} \int \limits_{-\infty}^{+\infty} d\eta \int
\limits_{t-r}^{t+r} d\xi \, \frac {P_1(\mu)} {(\xi-\eta)^2}  \left[ \frac {d} {d \eta} \left(
I^{1}_{2}(\eta) a''(\eta) \right) \right.
\nonumber\\
&-& \left. \frac{1}{\xi-\eta} \left( I^{1}_{2}(\eta) a'(\eta) - \frac {d} {d \eta} U_1(\eta)
\right)\, + \mathcal{O} \left( \frac{1} {(\xi-\eta)^2} \right) \right],
\end{eqnarray}
where
\begin{equation}\label{Ul}
U_1(\eta) = 4 I^{1}_{2}(\eta) a'(\eta) + (2 I^{2}_{2}(\eta) - 5 I^{0}_{1}(\eta)) a''(\eta)\,.
\end{equation}
Performing the inner integral over $\xi$ in (\ref{F3(3)}) with the help of the identity
(\ref{master}) we get the asymptotic behavior which is valid for large retarded times $u$
\begin{equation}
\label{F3tail} F_3(t,r) = \frac{r}{(t^2-r^2)^2} \left[\kappa A_1 + \mathcal{O}\left( \frac{1}
{t}\right)\right]\,,
\end{equation}
where
\begin{equation}
\label{A1} A_1 = \frac{8}{3} \int \limits_{-\infty}^{+\infty} \left(a''(s)\right)^2 a(s)\, ds\,.
\end{equation}
From (\ref{F3tail}) we obtain the late-time tails in both asymptotic regimes: $F_3(t,r)
\simeq\kappa A_1 r t^{-4}$ at future timelike infinity ($r=const,t\rightarrow\infty$) and $(r
F_3)(v=\infty,u) \simeq \kappa A_1 (2u)^{-2}$ at future null infinity
($v=\infty,u\rightarrow\infty$).
\subsection{$\mathbf{\ell\geq 2}$}
We give the detailed calculation only for $\ell=2$. In this case we have from (\ref{Beta2})
\begin{eqnarray}
\label{beta2_l2} \beta_2(u,r) &\stackrel{r+u>R}{=}& \frac {\kappa} {r^2} \left[ 2 I^{1}_{3}(u) +
\frac{1}{r} (2 I^{2}_{3}(u) - 13 I^{0}_{2}(u)) + \frac{1}{r^2} (2 I^{3}_{3}(u) - 39 I^{1}_{2}(u)
+ 30 \left(a'(u)\right)^2) + \mathcal{O} \left( \frac {1} {r^3} \right) \right]\,,
\\
\label{beta2dot_l2} \dot{\beta}_2(u,r) &\stackrel{r+u>R}{=}& - \frac {\kappa} {r^2} \left[ 2
I^{0}_{3}(u) + \frac{1}{r} (4 I^{1}_{3}(u) - 13 \left(a''(u)\right)^2) + \frac{1}{r^2} (6
I^{2}_{3}(u) - 39 I^{0}_{2}(u) - 60  a'(u) a''(u)) + \mathcal{O} \left( \frac {1} {r^3} \right)
\right] \,,
\\
\label{beta2prime(2)_l2} \beta_2'(u,r)  &\stackrel{r+u>R}{=}& \frac {\kappa} {r^2} \left[ 2
I^{0}_{3}(u) - \frac{13}{r} \left(a''(u)\right)^2 - \frac{60}{r^2}  a'(u) a''(u) + \mathcal{O}
\left( \frac {1} {r^3} \right) \right]\,.
\end{eqnarray}
Substituting (\ref{f1retadv}) and (\ref{beta2_l2}-\ref{beta2prime(2)_l2}) into (\ref{f3(2)}) we
obtain
\begin{eqnarray}
\label{F3(3)_l2} F_3(t,r) &=& \frac {4} {r} \int \limits_{-\infty}^{+\infty} d\eta \int
\limits_{t-r}^{t+r} d\xi \, \frac {P_l(\mu)} {(\xi-\eta)^2}  \left[ \kappa \frac {d} {d \eta}
\left( I^{1}_{3}(\eta) a^{(3)}(\eta) \right) + \frac{\kappa}{\xi-\eta} \left( -5 I^{1}_{3}(\eta)
a''(\eta) + \frac {d} {d \eta} U_2(\eta) \right) \right.
\nonumber\\
&+& \left. \frac{4}{(\xi-\eta)^2} \left( \left( a''(\eta) \right)^3 + 2 \kappa \left( 2 \left(
a''(\eta) \right)^3 - 3 \left(a^{(3)}(\eta)\right)^2 a(\eta) + \frac{1}{8} \frac {d} {d \eta}
V_2(\eta) \right) \right)\, + \mathcal{O} \left( \frac{1} {(\xi-\eta)^3} \right) \right],
\end{eqnarray}
where
\begin{equation}\label{U2}
U_2(\eta) = 8 I^{1}_{3}(\eta) a''(\eta) + (2 I^{2}_{3}(\eta) - 11 I^{0}_{2}(\eta))
a^{(3)}(\eta)\,,
\end{equation}
and
\begin{equation}\label{V2}
V_2(\eta) = -24 I^{0}_{3}(\eta) a(\eta) + 36 I^{1}_{3}(\eta) a'(\eta) + \left( -117
I^{0}_{2}(\eta) + 18 I^{2}_{3}(\eta) \right) a''(\eta) + \left( -78 I^{1}_{2}(\eta) + 4
I^{3}_{3}(\eta) + 60 \left( a'(\eta) \right)^2 \right) a^{(3)}(\eta)\,.
\end{equation}
Performing the inner integral over $\xi$ in (\ref{F3(3)_l2}) with the help of the identity
(\ref{master}) we get the asymptotic behavior for large retarded times
\begin{equation}
\label{F3tail_l2} F_3(t,r) = \frac{r^2}{(t^2-r^2)^3}\,\left[\kappa A_2 +B_2 + \mathcal{O}
\left(\frac{1}{t} \right)\right] \,,
\end{equation}
where
\begin{equation}
\label{B2} A_2=\frac{128}{15} \int \limits_{-\infty}^{+\infty} \left[ 2 \left(a''(s)\right)^3 - 3
\left(a^{(3)}(s)\right)^2 a(s) \right] ds \quad \mbox{and} \quad B_2 = \frac{64}{15} \int
\limits_{-\infty}^{+\infty} \left(a''(s)\right)^3  ds\,.
\end{equation}
For the general $\ell$ it is easy to see that the first nonzero contribution to the tail comes
from the term with $n=\ell+2$ in the identity (\ref{master}) which gives the following
 asymptotics
\begin{equation}\label{F3tail_lgen}
 F_3(t,r) = \frac{r^{\ell}}{(t^2-r^2)^{\ell+1}}\,\left[\kappa A_{\ell} +B_{\ell} + \mathcal{O}
\left(\frac{1}{t} \right)\right]\,.
\end{equation}
 The formula
(\ref{F3tail_lgen}) gives the first term in the asymptotic series approximation of the solution
for late retarded times, that is for small $\varepsilon$ we have
\begin{equation}\label{}
    \frac{(t^2-r^2)^{\ell+1}}{r^{\ell}} |F(t,r)-\varepsilon^3 F_3(t,r)| =
    \mathcal{O}(\varepsilon^5)\,.
\end{equation}
We have not attempted to derive a general formula for the coefficients $A_{\ell}$ and $B_{\ell}$
-- the computation of these coefficients for each given $\ell$ is straightforward but as $\ell$
increases the algebra becomes tedious  since it involves high-order expansions of the metric
functions along the light cone. Anyway, it follows from (\ref{F3tail_lgen}) that the tail behaves
as $F_3(t,r)\sim r^{\ell} t^{-(2\ell+2)}$ at future timelike infinity and as $(r
F_3)(v=\infty,u)\sim u^{-(\ell+1)}$ at future null infinity.

\vskip 0.2cm \noindent \emph{Remark 1.} For $\ell\geq 2$ the tail (\ref{F3tail_lgen}) has two
parts quantified by the coefficients $\kappa A_{\ell}$ and $B_{\ell}$, respectively. The
$A_{\ell}$-part comes from the gravitational self-interaction of the wave map and vanishes for
$\kappa=0$. The $B_{\ell}$-part comes from the cubic nonlinearity of the wave map equation and is
present without gravity as well. The case $\ell=1$ is special in the sense that the
$B_{\ell}$-part is absent in (\ref{F3tail}) since it is subdominant (decaying as $t^{-5}$) with
respect to the leading order term. \vskip 0.2cm \noindent
 \emph{Remark 2.} It is instructive to compare the tail (\ref{F3tail_lgen})
with the tail for a test linear massless field propagating on a fixed stationary asymptotically
flat background. According to the Price law \cite{price,gpp1,poisson} the $\ell$-th multipole of
this linear tail  $\phi_{\ell}(t,r)\sim r^{\ell+1}/(t^2-r^2)^{\ell+2}$ for
$t-r\rightarrow\infty$. This decay is by
 one power faster than that in (\ref{F3tail_lgen}). Of course, this difference is not very surprising
 as the tail
 studied here and Price's tail correspond to  different physical situations, however we point it out
 as another example of the inapplicability of linearized theory in the study of radiative relaxation
 processes (see \cite{BCR3,BCR2} for other examples).
  We shall discuss this issue in more detail elsewhere
 \cite{br}.
\section{Numerics}
\label{sec:numerics} In this section we compare the above analytic predictions with  the results
of numerical solutions of Einstein-wave map equations (\ref{h-constraint}-\ref{we2}) for various
initial data. The details of the numerical method were given in \cite{bcr5} for the case
$\ell=0$. The only difference for higher $\ell$ is the boundary condition $F(t,r)\sim r^{\ell}$
for small $r$ which guarantees regularity at the origin.
The initial data were generated by the gaussian
\begin{equation}\label{idn}
\varepsilon a(x) = \varepsilon  \exp\left(-x^2\right)
\end{equation}
for different values of $\varepsilon$. For these initial data the formula (37) gives for $\ell=1$
\begin{equation}\label{coeff_l1}
A_1=\frac{64}{27}\sqrt{3\pi}\approx 7.2769\,,
\end{equation}
and the formula (45) gives for $\ell=2$
\begin{equation}\label{coeff_l2}
A_2=-\frac{10240}{81}\sqrt{3\pi}\approx -388.1\,,\qquad B_2=-\frac{2048}{405}\sqrt{3\pi}\approx
-15.52\,.
\end{equation}
In order to extract the parameters of the tails at timelike infinity we fit our numerical data
with the formula
\begin{equation}\label{tail_fit}
 F(t,r) = A t^{\gamma} \exp \left(B/t + C/t^2 \right)\,.
\end{equation}
 \begin{figure}[h]
 \vskip -0.2cm
\begin{tabular}{cc}
\includegraphics[width=0.45\textwidth]{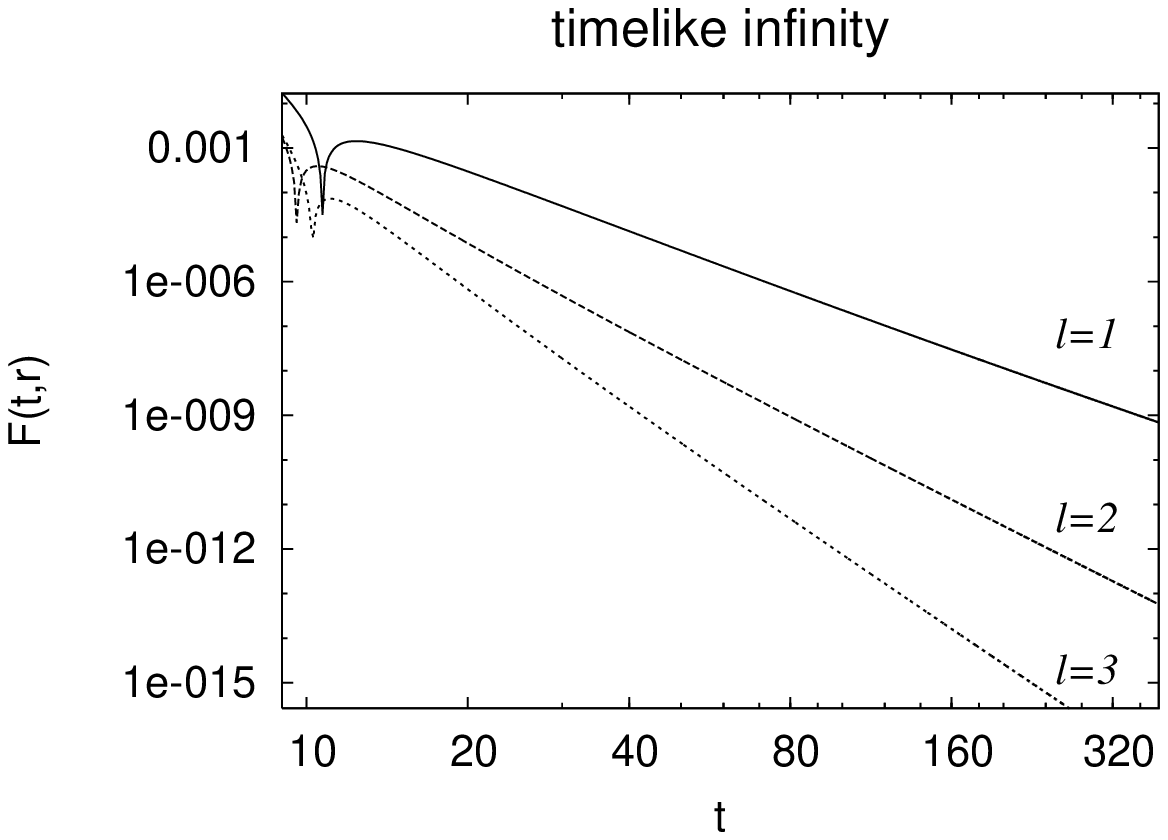}
&
\includegraphics[width=0.45\textwidth]{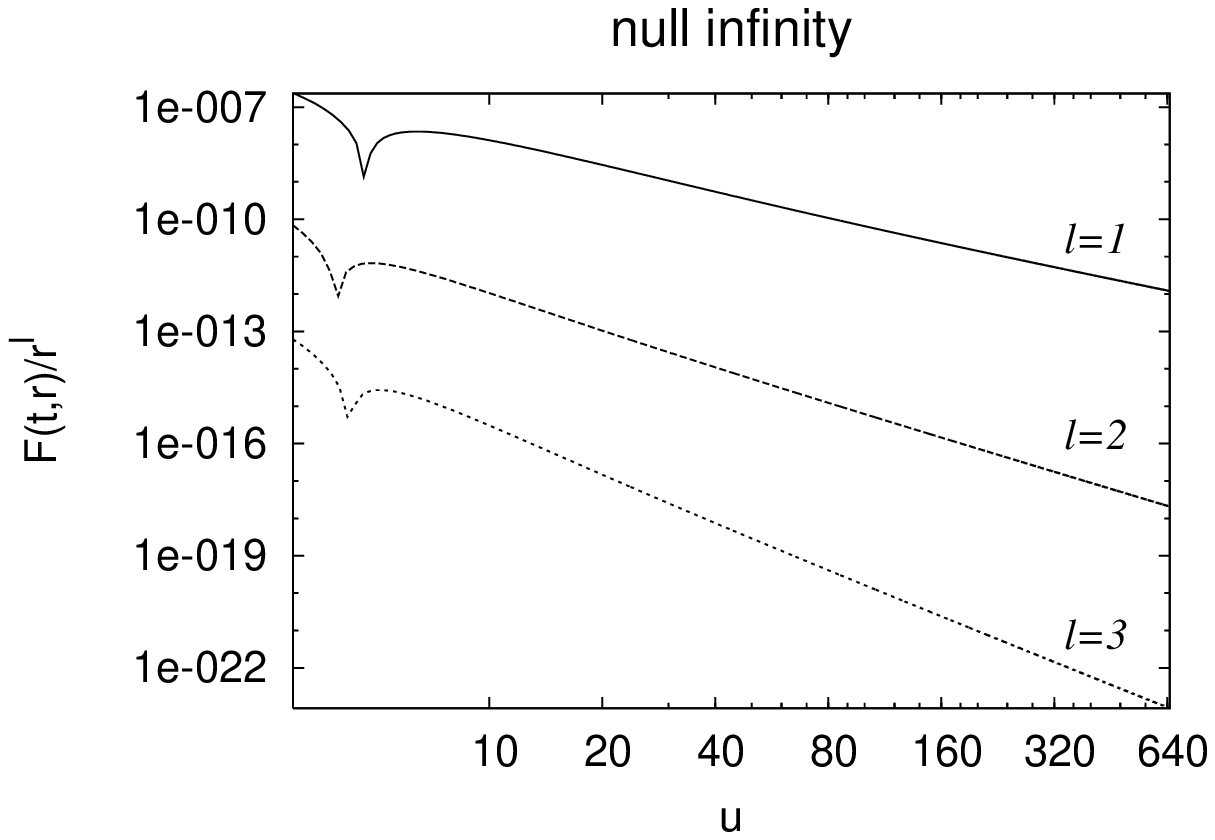}
\\
\end{tabular}
\caption{\small{Left panel: The log-log plot of $F(t,r)$ for fixed $r=5$. Fitting
(\ref{tail_fit}) we get power-law exponents $\gamma=-4.0196$ ($\ell=1$), $-6.0009$ ($\ell=2$),
$-8.0049$ ($\ell=3$), in agreement with the analytic prediction (46). Right panel: The log-log
plot of $F(t,r)/r^{\ell}$ for fixed large advanced time $v=t+r=1200$ as the function of retarded
time $u=t-r$. The analogous fit to (51) yields the exponents $-2.0036$ ($\ell=1$), $-3.0004$
($\ell=2$), $-4.0095$ ($\ell=3$), in accordance with (46). In both panels $\kappa=0.02$ and
$\varepsilon=2.0$ ($\ell=1)$, $\varepsilon=0.7$ ($\ell=2)$, $\varepsilon=0.3$ ($\ell=3)$. }}
\label{fig.null}
\end{figure}

\begin{table}[h]
\begin{center}
\begin{tabular}
{|c|c|c||c|c|c|} \hline  & \multicolumn{2}{c||}{$A(\ell=1)$}  &  &
\multicolumn{2}{c|}{$A(\ell=2$)} \\
\cline{2-3} \cline{5-6}
$\varepsilon$ & theory & numerics & $\varepsilon$  & theory & numerics \\
\hline\hline
$\,\,\,$0.05$\,\,\,$ &$\,\,\,$ 9.096e-5 $\,\,\,$& $\,\,\,$9.051e-5$\,\,\,$ & $\,\,\,$0.05 $\,\,\,$
& $\,\,\,$-0.07277$\,\,\,$ & $\,\,\,$-0.07274$\,\,\,$\\
\hline
0.1 & 7.277e-4 & 7.289e-4 & 0.1 & -0.58216 & -0.58476 \\
\hline
0.4 & 0.04657 & 0.04701 & 0.2 & -4.65727 & -4.6841\\
\hline
0.8 & 0.37258 & 0.37414 & 0.4 & -37.2582 & -37.3778\\
\hline
2.4 & 10.0597 & 10.0299 & 0.65 & -159.875 & -160.441 \\
\hline
3.2 & 23.8452 & 16.0528 & 0.7 & -199.681 & -189.377  \\
\hline
3.8 & 39.9303 & 19.6931 & 0.75 & -245.598 & -189.792 \\
\hline
\end{tabular}
\end{center}
\caption{The comparison of analytic and numerical amplitudes of the tails at timelike infinity.
Here $\kappa=0.02$ and $r=5$. \\The third-order approximation reads $A=\varepsilon^3 \kappa r
A_1$ for $\ell=1$ and  $A=\varepsilon^3 r^2(\kappa A_2+B_2)$ for $\ell=2$.}
\end{table}

\begin{figure}[h]
\begin{tabular}{cc}
\includegraphics[width=0.45\textwidth]{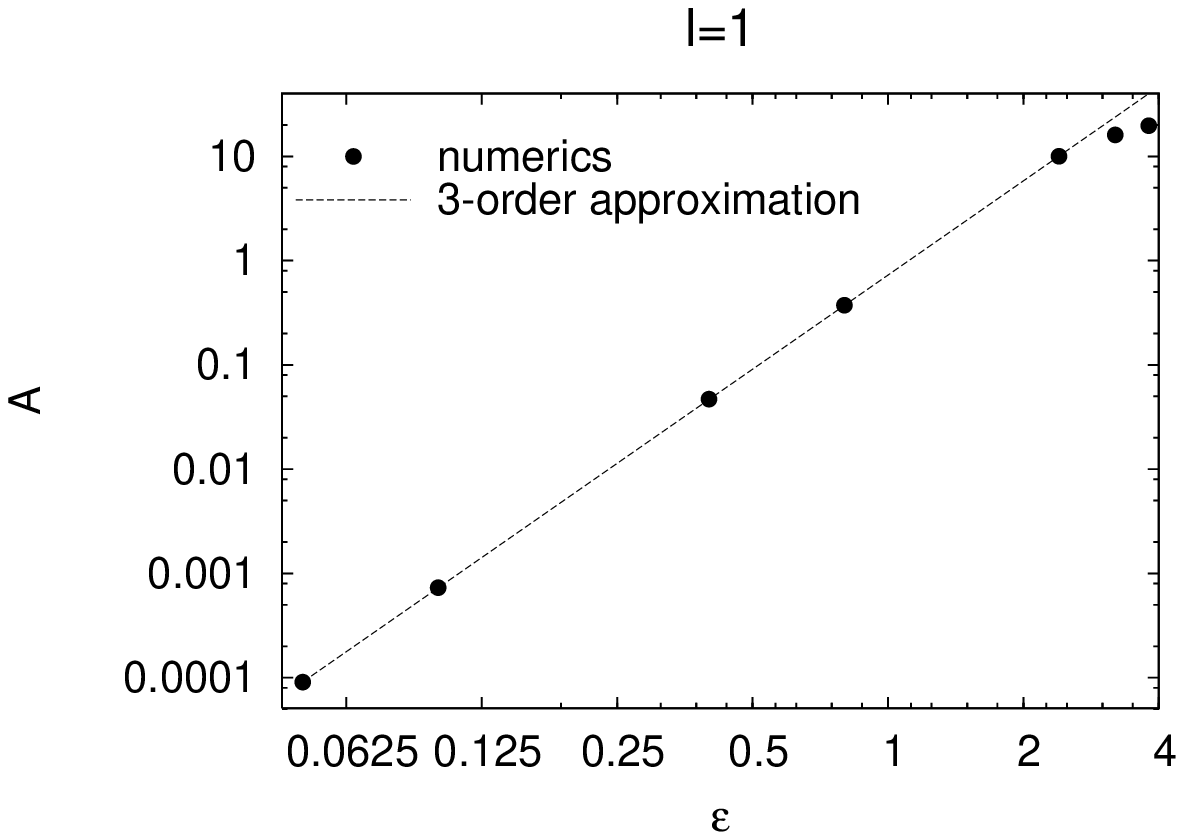}
&
\includegraphics[width=0.45\textwidth]{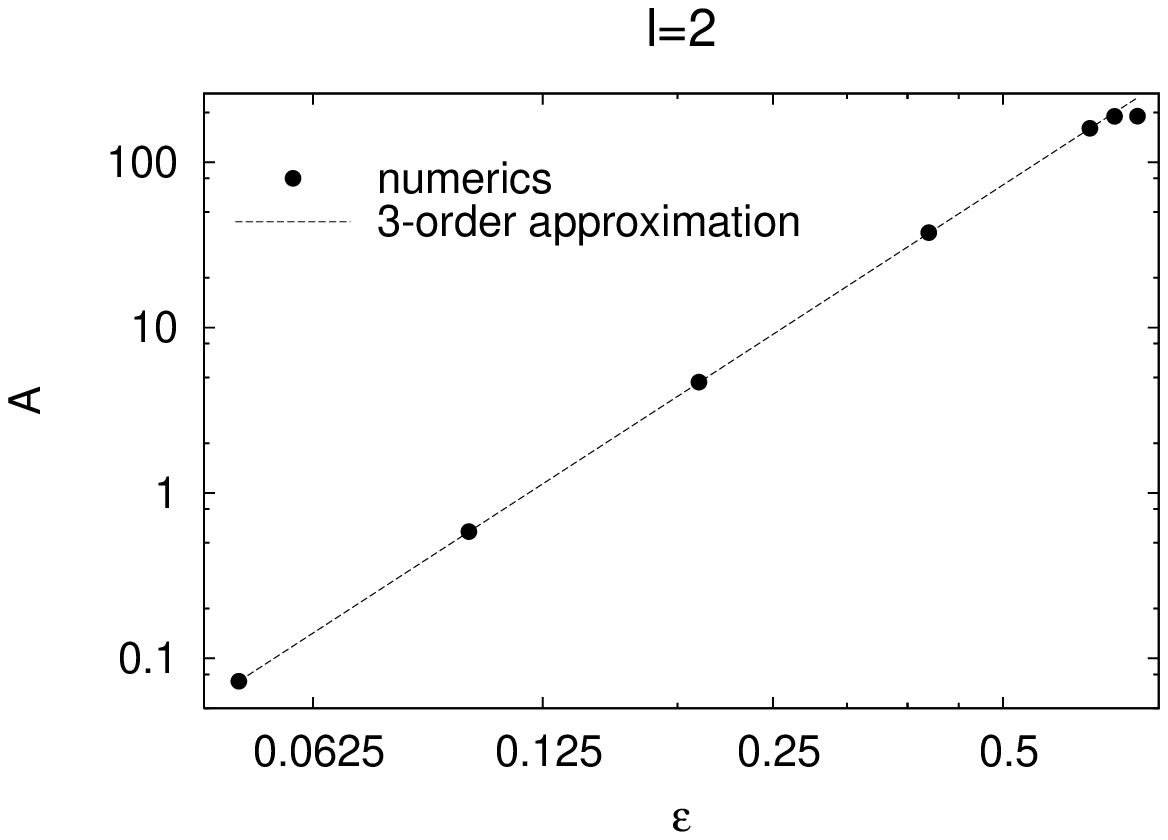}
\\
\end{tabular}
\caption{\small{The log-log plot of the amplitude of the tail at timelike infinity as a function
of the amplitude of initial data (black dots) for fixed $\kappa=0.02$ and $r=5$. The third-order
approximation (dashed line) is excellent for small data, but it breaks down for large data lying
near the threshold of black hole formation.}} \label{fig.ampl}
\end{figure}
\begin{figure}[h]
\begin{tabular}{cc}
\includegraphics[width=0.45\textwidth]{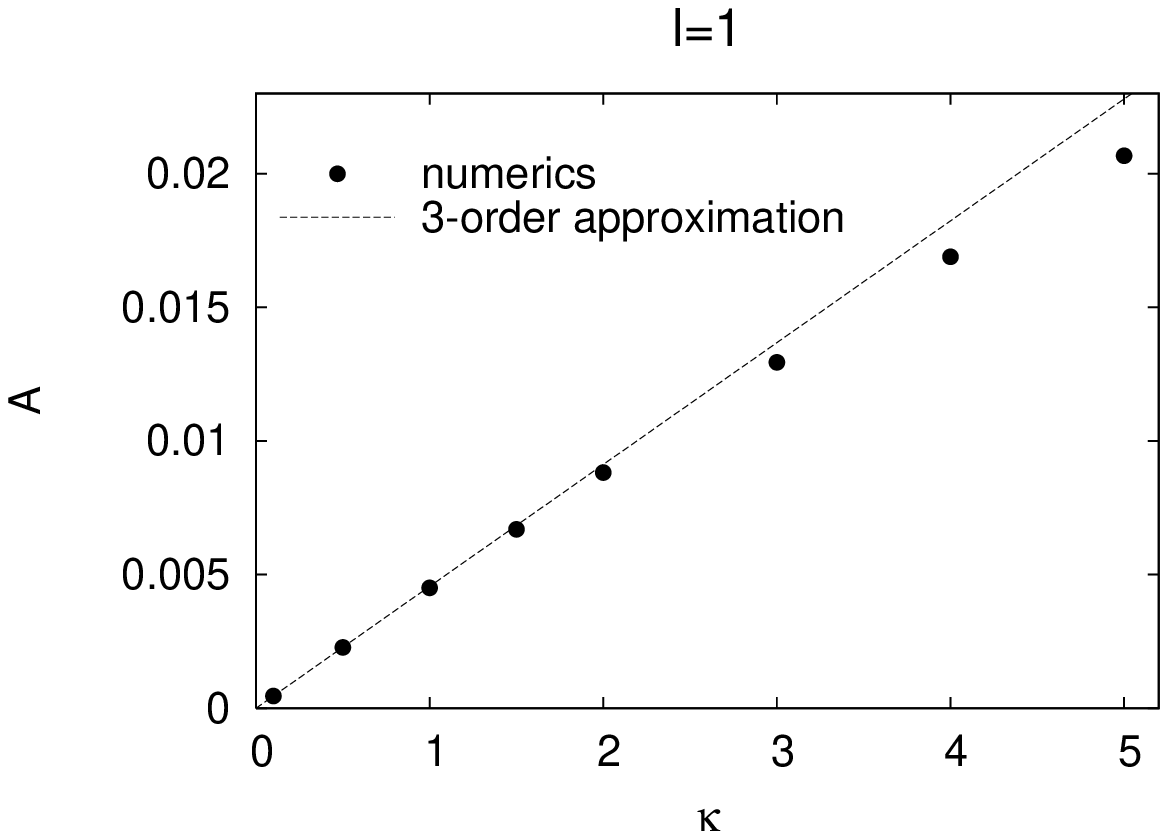}
&
\includegraphics[width=0.45\textwidth]{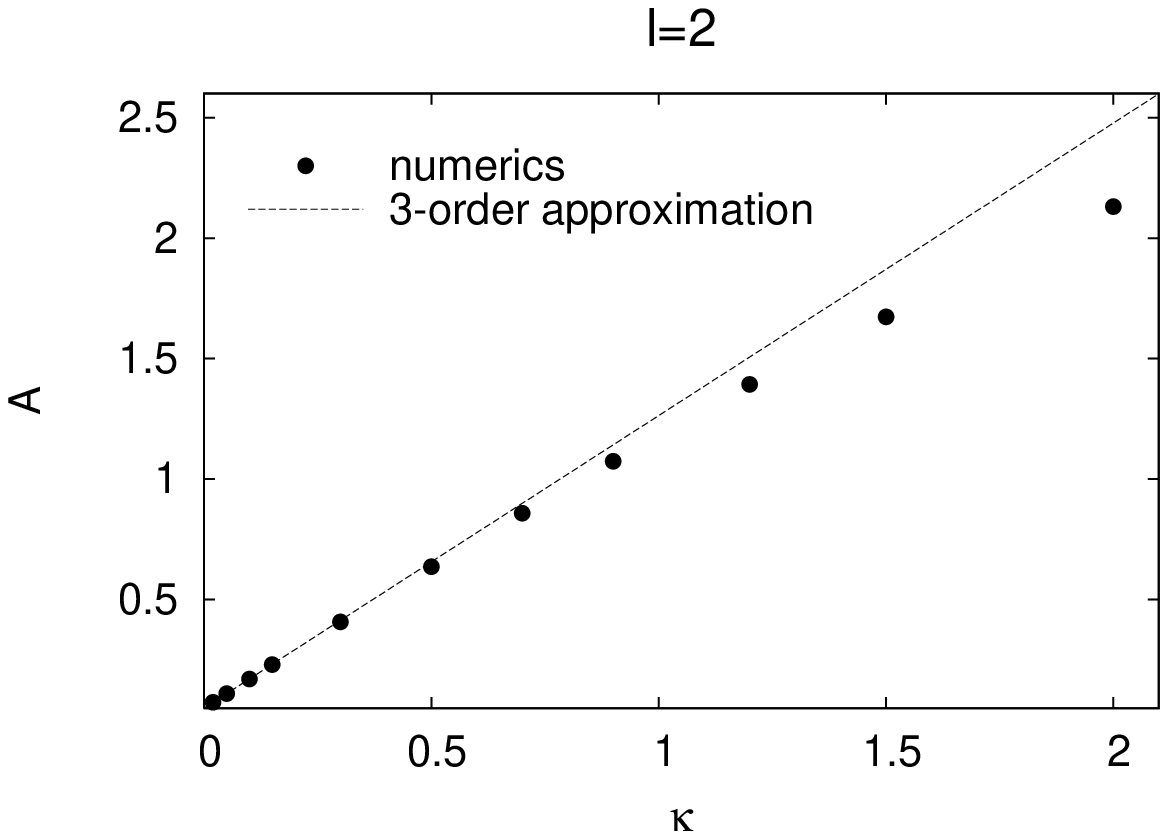}
\\
\end{tabular}
\caption{\small{The plot of the amplitude of the tail at  timelike infinity as a function of the
coupling constant $\kappa$ (black dots) for fixed $\varepsilon=0.05$ and $r=5$. As $\kappa$
increases we leave the small-data regime and consequently the third-order approximation (dashed
line) deteriorates.}} \label{fig3}
\end{figure}
 The results  and their confrontation with analytic predictions are
summarized in Table~1 and Figures~1, 2, and 3.  From this comparison we conclude that the
third-order approximation
 is excellent for sufficiently small initial data. For large data approaching the black-hole
 threshold the third-order approximation breaks down -- this is seen in Fig.~2 as the deviation
  from the scaling
 $A\sim \varepsilon^3$ and in Fig.~3 as the deviation from the linear dependence of $A$ on
 $\kappa$.\\ It should be emphasized that we get the same
  decay rates $t^{-(2\ell+2)}$ (at timelike infinity) and $u^{-(\ell+1)}$ (at null infinity) for
\emph{all} subcritical evolutions, regardless of whether our third-order formula reproduces
accurately the amplitude of the tail (for small data) or fails (for large data).

 \vskip 0.2cm \noindent \textbf{Acknowledgments:} We acknowledge
support by the MNII grants: NN202 079235 and 189/6.PRUE/2007/7.

\end{document}